\documentclass[final,5p,times,twocolumn]{elsarticle}
\journal{Physics Letters B}

\usepackage{graphicx} 
\usepackage{subfigure}
\usepackage{amssymb}
\usepackage{units}
\usepackage{xcolor}
\usepackage{amsmath}
\usepackage{units}
\biboptions{compress}
\usepackage{hyperref}

\newcommand{\Fig}[1]{{Fig.~\ref{#1}}}
\newcommand{\taups}{$\tau_{\text{PS}}$}

\begin{document}

\begin{frontmatter}

\title{Jet Quenching Without Energy Loss}

\author[add1,add2]{Liliana Apolinário}
\author[add3]{Chiara Le Roux}
\author[add3]{Korinna Zapp}
\affiliation[add1]{organization={LIP},
            addressline={Av. Prof. Gama Pinto 2}, 
            city={Lisboa},
            postcode={P-1649-003}, 
            state={},
            country={Portugal}}
\affiliation[add2]{organization={Departamento de  Física, Instituto Superior Técnico (IST), Universidade de Lisboa},
            addressline={Av. Rovisco Pais 1}, 
            city={Lisboa},
            postcode={P-1049-001}, 
            state={},
            country={Portugal}}
\affiliation[add3]{organization={Dept. of Physics, Lund University},
            addressline={Sölvegatan 14A}, 
            city={Lund},
            postcode={S22362}, 
            state={},
            country={Sweden}}

\begin{abstract}

The onset of jet quenching, i.e. the suppression of high transverse momentum particles and jets, is an important question in the context of understanding the onset of collective behaviour and small collision systems. We investigate a minimal scenario where a hard parton experiences a single soft re-scattering that leaves the kinematics unmodified, but the colour exchange leads to a loss of colour coherence that is observable in the final distribution of fragments. In particular, the formation time of the first splitting as reconstructed at hadron level from jets using the formation time clustering algorithm is sensitive the loss of colour coherence. Moreover, it can distinguish the coherence loss scenario from a corresponding energy loss scenario, since small energy loss effects leave the formation time distribution unchanged.

\end{abstract}


\end{frontmatter}

\section{Introduction}

More than a decade after the discovery of signs of collective behaviour in small collision systems there are still many open questions regarding the dynamics behind these observations (see~\cite{Dusling:2015gta,Song:2017wtw,Nagle:2018nvi,Apolinario:2022vzg, Noronha:2024dtq,Grosse-Oetringhaus:2024bwr} for reviews). In particular, the apparent absence of suppression of high transverse momentum particles and jets, known as jet quenching, challenges the standard picture of heavy ion collisions in which jet quenching and for instance azimuthal anisotropy have a common origin in final state re-scattering. It is therefore an important task to identify the onset of jet quenching. This is an experimental as well as a theoretical challenge because it requires predictions for how small jet modifications could manifest themselves in data. 

\smallskip

It was realised in~\cite{Mehtar-Tani:2010ebp,Mehtar-Tani:2011hma,Casalderrey-Solana:2011ule} that re-scattering in a dense, coloured background medium leads to a loss of colour coherence and thus a loss of angular ordering in the parton shower\footnote{Coherence and loss of angular ordering was also discussed in a somewhat different context in~\cite{Caucal:2018dla} and implemented in~\cite{Caucal:2019uvr}}. We here argue that this could be the first observables sign of the onset of jet modification.

In a weak coupling scenario where jet quenching is due to scattering of hard partons off a coloured background already the first scattering, even if it is too soft to lead to any sizable energy loss, changes the colour of the hard parton. If this happens early, i.e.\ during the parton shower evolution, the change in colour destroys the colour coherence of the hard partons and leads to a loss of angular ordering. It therefore changes the phase space available for radiation off the hard parton, which leads to modifications of the (potentially hard) radiation pattern. The goal of the present work is to investigate how the loss of colour coherence could be observed experimentally, and to contrast it with a scenario in which already the onset of jet quenching is a kinematical effect coming from energy loss of hard partons. In particular, it is shown that the formation time of parton shower emissions as estimated using the $\tau$ clustering algorithm~\cite{Apolinario:2020uvt} is sensitive to angular ordering. Moreover, the formation time is able to distinguish the coherence loss and the energy loss scenario. These results, obtained within a controlled Monte Carlo event-generator setup and based on a specific model implementation, mostly rely on an explicit breaking of colour coherence, thus ensuring their robustness. They highlight the significant physics potential of upcoming lighter-ion runs at the LHC to uncover signatures of QGP droplets that would otherwise remain difficult to access experimentally. 

\section{The model}
\label{sec:model}

We here consider a minimal scenario in which a re-scattering occurring before the first splitting of the parton shower leads to a loss of angular ordering for the first splitting, but leaves the kinematics completely unmodified. This amounts to running the QCD evolution as in vacuum, but with angular ordering disabled (only) for the first emission. This is motivated by small collision systems, where one would generically expect few final state re-scatterings taking place at an early time, because theses systems are small and short-lived. Events are generated with \textsc{Jewel}~\cite{Zapp:2012ak,Zapp:2013vla}, which is based on \textsc{Pythia}\,6.4~\cite{Sjostrand:2006za}, in the latest version 2.6.0\footnote{The code will soon be available at \url{jewel.hepforge.org}.}~\cite{ccpaper} that allows for a more flexible handling of angular ordering. Re-scattering is not simulated explicitly, thus only the vacuum evolution is used. For angular ordering we run the option where kinematic constraints are imposed before the angular ordering condition (cf.~\cite{ccpaper}). In the alternative energy loss scenario angular ordering is required for all splittings, but the momenta of the partons coming out of the hard scattering matrix element are reduced by a constant fraction to mimic energy loss. The momenta are modified after the scale of the first splitting has been found and the kinematics (with off-shell partons) has been determined. We run a standard setup with initial state radiation and hadronisation, but without multi-parton interactions.

The events are analysed using Rivet~\cite{Bierlich:2019rhm}, HepMC~\cite{Buckley:2019xhk}, FastJet~\cite{Cacciari:2011ma} and \textsc{Yoda}. For the results presented in section \ref{sec:results}, we have generated $Z+$Jet events with center-of-mass energy $\sqrt{s}=\unit[5.02]{TeV}$. In the analysis of the simulation results, we use all particles in a rapidity interval $|\eta_{\text{cut}}| \le 2.8$ with $p_\perp \ge \unit[0.5]{GeV}$. We first reconstruct the $Z$-boson via the di-muon channel requiring an invariant mass between 80 and \unit[100]{GeV} and a minimum transverse momentum $p_{\perp,\text{min}}=\unit[100]{GeV}$. The muons are also required to have $p_\perp \ge \unit[5]{GeV}$. Once the $Z$ is reconstructed, the remaining particles are clustered into jets using the anti-$k_\perp$ algorithm~\cite{Cacciari:2008gp}. The jet radius used was $R=1.0$ to ensure a reasonably complete reconstruction of the fragments. The minimum jet $p_\perp$ is set to \unit[25]{GeV} and, in order to avoid border effects, they are required to be within a rapidity interval of $|\eta_{\text{cut}}-R|$. The $Z$-boson and the jets are then required to have an azimuthal separation of at least $\Delta \phi_{\text{min}}=5\pi/8$, and among these, the hardest jet is selected for further analysis. With these requirements, the probability for finding a jet in events containing a $Z$ boson is $99.7$\% in all the set-ups implemented. Finally, once the event has passed all the selection criteria, the hardest jet is reclustered with the $\tau-$algorithm with a radius $R=1.5$. The jet is then unclustered and the two sub-jets found after the first unclustering step are taken to be the two prongs for the calculation of the formation time. Following~\cite{Apolinario:2020uvt} the formation time as estimated using
\begin{equation}
    \label{eq:formationtime}
    \tau=\frac{1}{2Ez(1-z)(1-\cos\theta_{12})},
\end{equation}
where $E$ is the energy of the jet, $z$ is the ratio of the energy of the soft subjet to the hard one, and $\theta_{12}={M}/{E \sqrt{z(1-z)}}$, with $M$ being the jet mass, is the angle between the two prongs.

It is important to emphasize that, in contrast to Refs.\cite{Apolinario:2020uvt,Apolinario:2024hsm}, no Soft Drop procedure is applied in the present analysis. In these earlier studies, Soft Drop grooming was introduced to enhance the correlation between the quantities extracted from jet re-clustering and those at the Monte Carlo truth level. Namely, in Ref.\cite{Apolinario:2020uvt}, where a jet radius of R = 0.4 was used, it was found that the main source of mismatch originated from wide-angle, soft emissions that were not captured within the jet radius and therefore did not contribute to the reconstructed jet. As a consequence, the first emission in the parton shower could not be associated with the first unclustering step of the reconstructed jet. This led to a subset of events where the two quantities were uncorrelated, predominantly at very small formation times in the parton shower, while the corresponding reclustered values were distributed over larger formation-time scales, reflecting that the first unclustering step corresponded to a later emission that happened to be captured within the jet. Applying Soft Drop both at the parton-shower and re-clustering levels improved this correspondence, as the wide-angle emissions responsible for the decorrelation were predominantly soft. 

In the present analysis, our focus is specifically on the first splitting and the goal is to retain all radiation relevant for potential large-angle coherence breaking. Since we use a larger jet radius, we therefore do not apply Soft Drop grooming to reduce the risk of not tagging the first splitting. We also note that, in this context, the Cambridge/Aachen algorithm would not provide a meaningful first unclustering step, since, without Soft Drop, it does not hold a physical interpretation. The anti-$k_T$ + $\tau$-reclustering combination thus provides the physically interpretable configuration for our purposes.

\section{Results}
\label{sec:results}

\begin{figure}[!ht]
    \centering
    \subfigure[]{
    \includegraphics[width=0.8\linewidth]{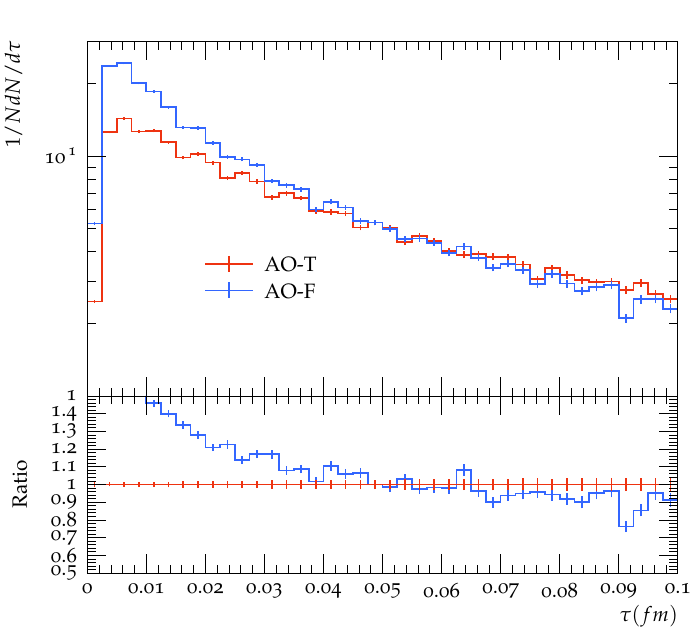}
    \label{fig:taudist_PS}
    } \\
    \subfigure[]{
    \includegraphics[width=0.8\linewidth]{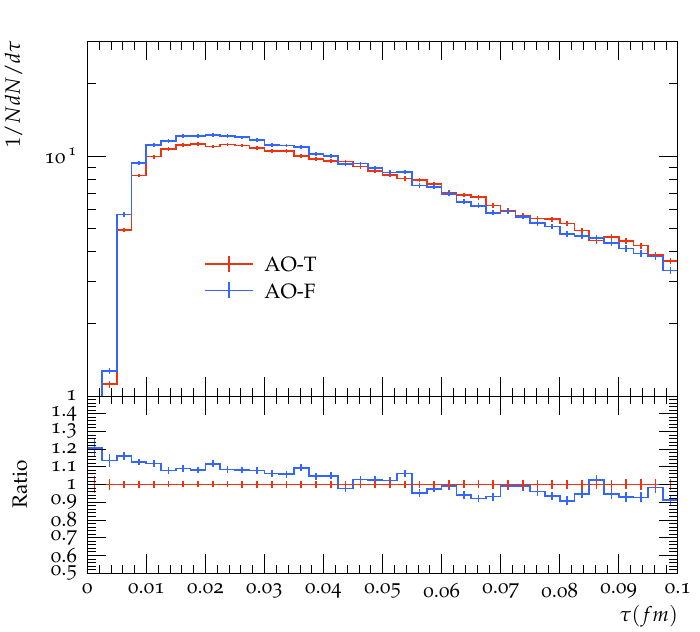}
    \label{fig:taudist_REC}
    }
    \label{fig:taudists}
    \caption{Formation time distribution of jets with radius $R=1.0$ for events with and without angular ordering at the first splitting. \subref{fig:taudist_PS} Formation time computed at the parton shower level. \subref{fig:taudist_REC} Formation time computed at the hadron level using the tau-algorithm unclustering procedure.}
\end{figure}

Using equation \ref{eq:formationtime} to compute the formation time at the level of the parton shower for events with a reconstructed $Z$ and a jet passing the cuts detailed above, we obtain the distribution in \Fig{fig:taudist_PS}. In this case, the two prongs correspond to the two partons into which the initial outgoing parton splits. The corresponding distribution obtained from the $\tau-$algorithm is shown in \Fig{fig:taudist_REC}. In each of the figures, two curves are shown: a solid blue line representing the setting in which the angle for the first splitting in the parton shower is not constrained, i.e., angular ordering is switched off for that splitting (AO-F) and a solid red line referring to the scenario in which the first splitting angle is constrained by the angle between the splitting parton and its final state color partner (AO-T).

It can be seen that, at parton shower level (Fig.~\ref{fig:taudist_PS}), there is a substantial modification of the formation time distribution between the two configurations. This difference is caused by a more limited phase space for the parton to split when angular ordering is enforced: smaller emission angles lead to longer formation times. Conversely, when angular ordering is disabled, the first emission can occur at larger angles, resulting in shorter formation times on average. This indicates that the modification of the formation time distribution can be a direct signal of the breaking of color coherence. When the formation time is estimated from reconstructed jets (\Fig{fig:taudist_REC}), the signal is somewhat attenuated, but it is still visible.

The differences between the two plots in \Fig{fig:taudists} are largely due to misreconstruction effects. Either due to the first emission ending up outside of the jet or to a mismatch between the parton shower and the unclustering histories.

\begin{figure}[!ht]
    \centering
    \subfigure[]{
    \includegraphics[width=0.8\linewidth]{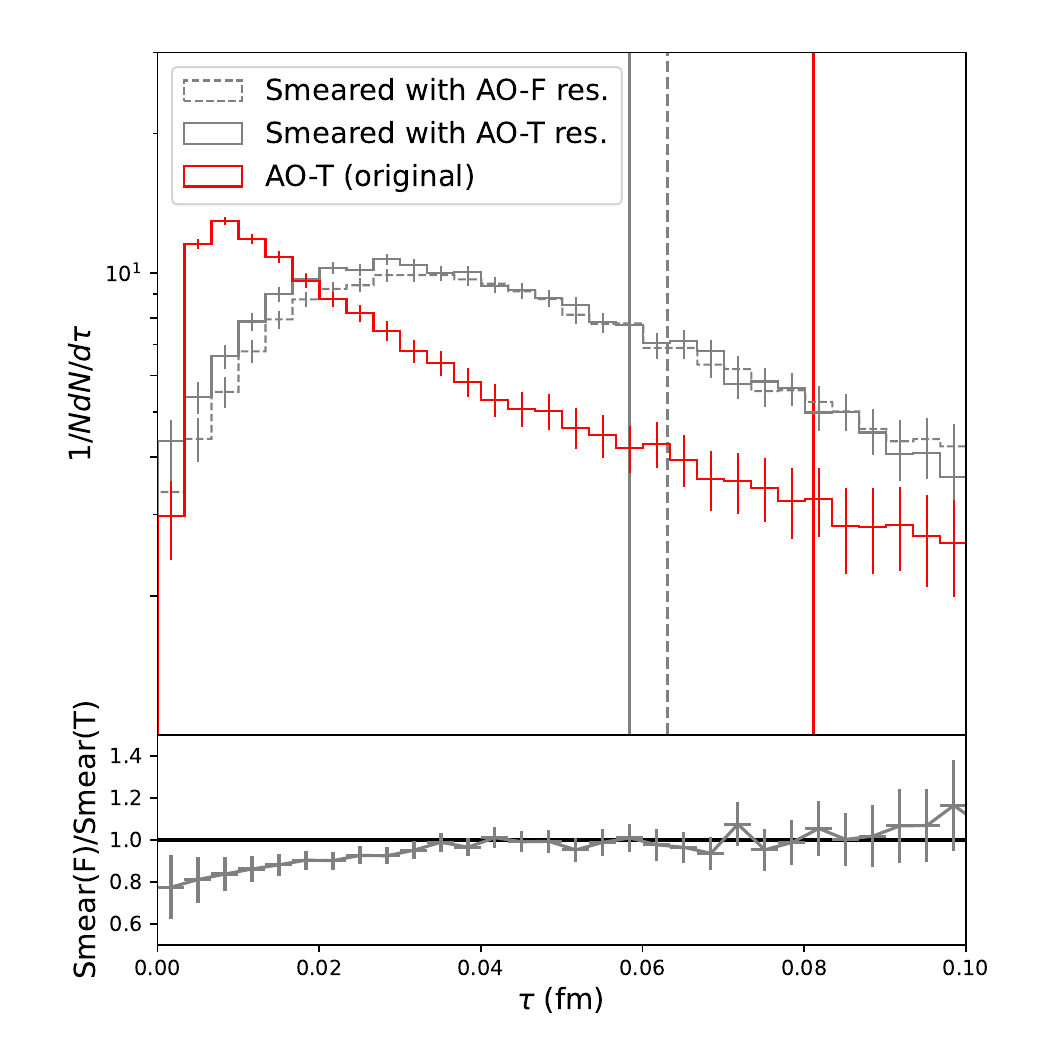}
    \label{fig:tausmearing-T}
    } \\
    \subfigure[]{
    \includegraphics[width=0.8\linewidth]{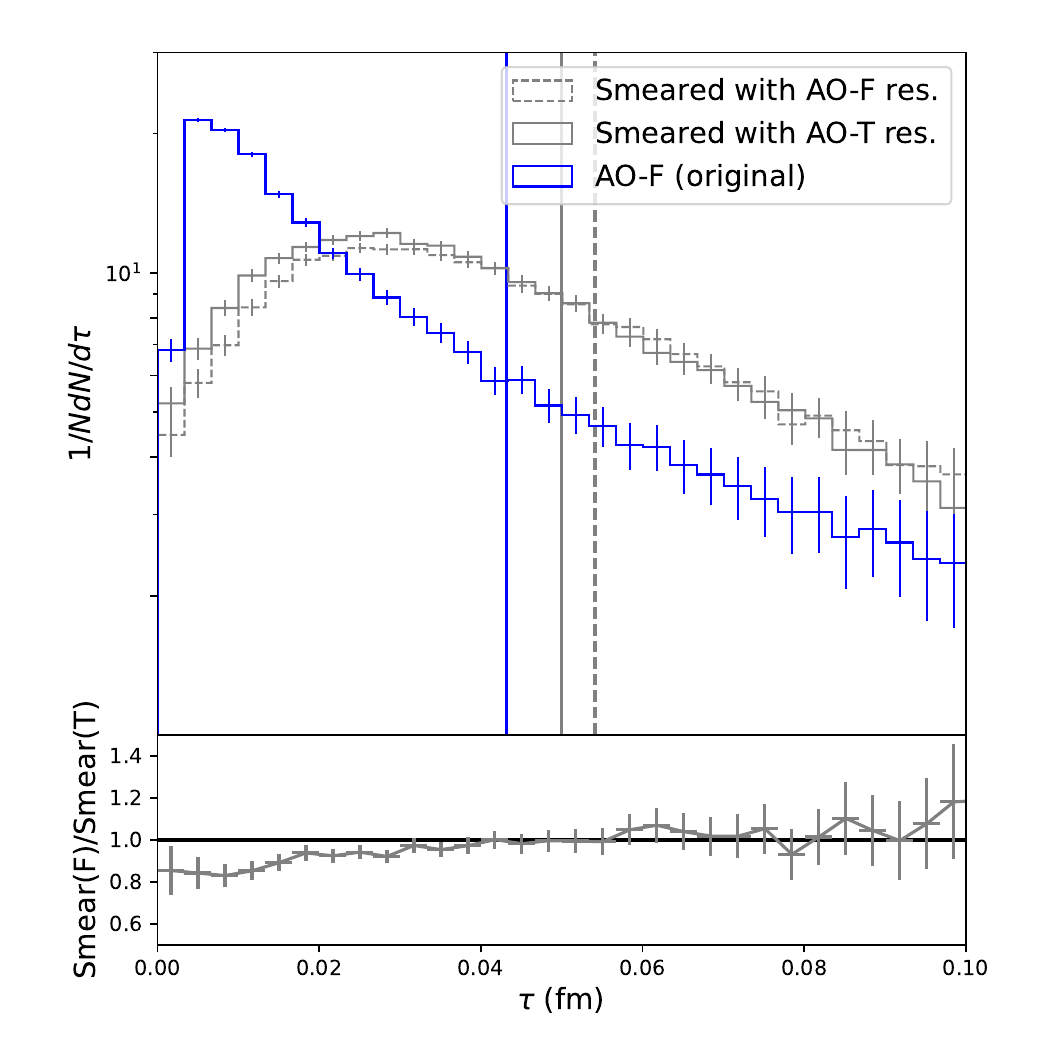}
    \label{fig:tausmearing-F}
    }
    \label{fig:tausmearing}
    \caption{The original \taups distribution for the \subref{fig:tausmearing-T} AO-T and the \subref{fig:tausmearing-F} AO-F configuration along with the corresponding smeared distributions. Solid grey lines show the smeared distribution resulting from the AO-T reconstruction resolution whereas dashed grey lines correspond to smearing from the AO-F resolution. The vertical lines show the value of the median of the corresponding distributions.}
\end{figure}

Due to these two effects the formation time is reconstructed with a finite resolution, which is slightly different for the cases with and without angular ordering for the first splitting. In order to better understand the effects of finite resolution and to verify that the difference in the formation time distributions observed at hadron level are not induced by the different resolution we smeared the same parton shower formation time distribution with the different resolutions. To this end we compute $\Delta\tau = \tau_{\mathrm{PS}} - \tau_{\mathrm{rec}}$ and determine its distribution in bins of $\tau_{\mathrm{PS}}$ for both the AO-T and AO-F configurations. Using these distributions as a reconstruction resolution, we then smear the original $\tau_{\mathrm{PS}}$ spectra by randomly sampling $\Delta\tau$ values according to the corresponding resolution in each bin. This means, of course, that smearing the original AO-T distribution with the resolution obtained from the AO-T reconstruction yields the same histogram as in the red line in \Fig{fig:taudist_REC}, and analogously for AO-F. Comparing the results of smearing either parton shower formation time distribution with the two different resolutions in \Fig{fig:tausmearing} it becomes evident that the different resolutions cannot be the origin of the effect observed in \Fig{fig:taudist_REC}. The effect of different resolution is actually opposite to that of different parton shower distributions: smearing with the resolution of the scenario without angular ordering leads to a larger observed formation time (compared to the case where we smear with the resolution of the scenario with angular ordering), while the reconstructed formation time without angular ordering is smaller than with angular ordering when the change in the underlying parton shower distribution is included. Therefore, we conclude that the difference observed in \Fig{fig:taudist_REC} is a result of the physical change in the distribution seen in \Fig{fig:taudist_PS}, and not an artifact of the reconstruction algorithm.

\begin{figure}[!ht]
    \centering
    \includegraphics[width=0.8\linewidth]{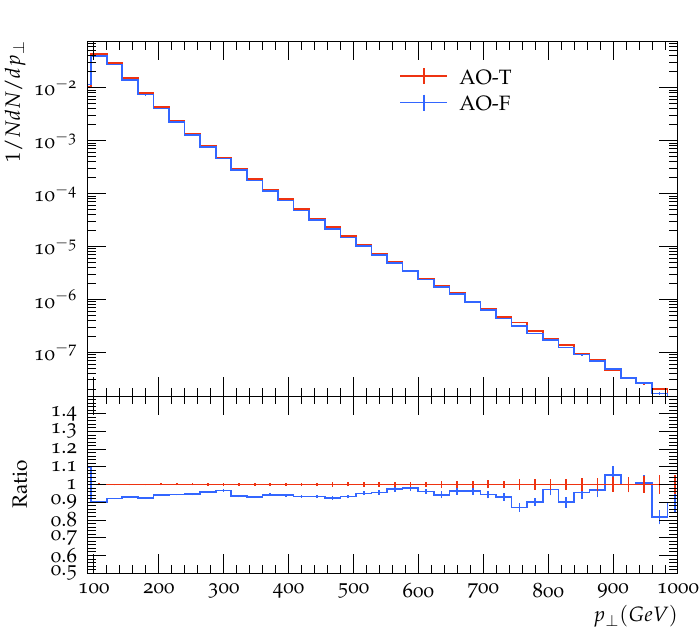}
    \label{fig:raa}
    \caption{Jet $p_\perp$ spectra for events with and without angular ordering in the first splitting for jets of radius $R=1.0$.}
\end{figure}

It should be emphasized that there is no parton energy loss in this scenario. The substantial effect shown in \Fig{fig:taudists} is due solely to turning angular ordering on or off in the first splitting of the parton shower. However, turning angular ordering off leads to  a broader distribution of fragments, which results in a smaller fraction of the initial parton's energy being reconstructed inside the jet. In that sense there is energy loss at the level of reconstructed jets. To quantify this we calculate the transverse momentum spectra and nuclear modification factor for the jets, shown in \Fig{fig:raa}. There is a suppression of around 5 to 10\% in the jet spectrum obtained for AO-F when compared to that with AO-T.

\begin{figure}[!ht]
    \centering
    \subfigure[]{
    \includegraphics[width=0.8\linewidth]{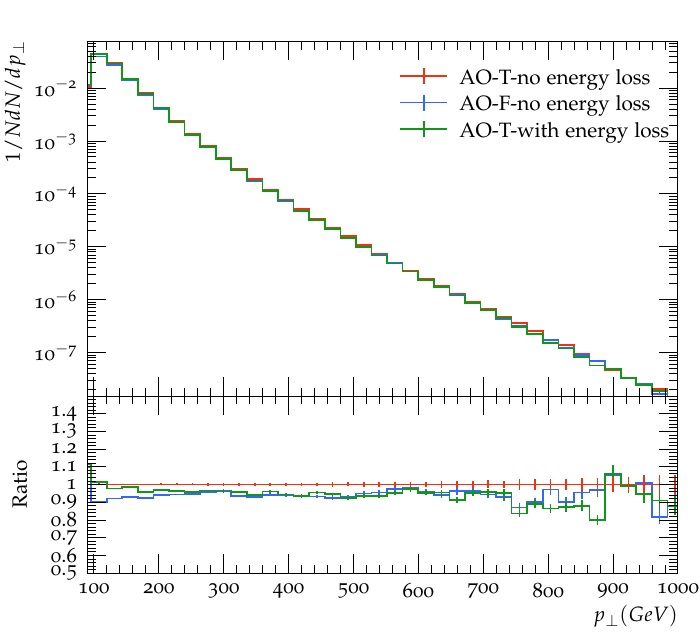}
    \label{fig:eloss-raa}
    } \\
    \subfigure[]{
    \includegraphics[width=0.8\linewidth]{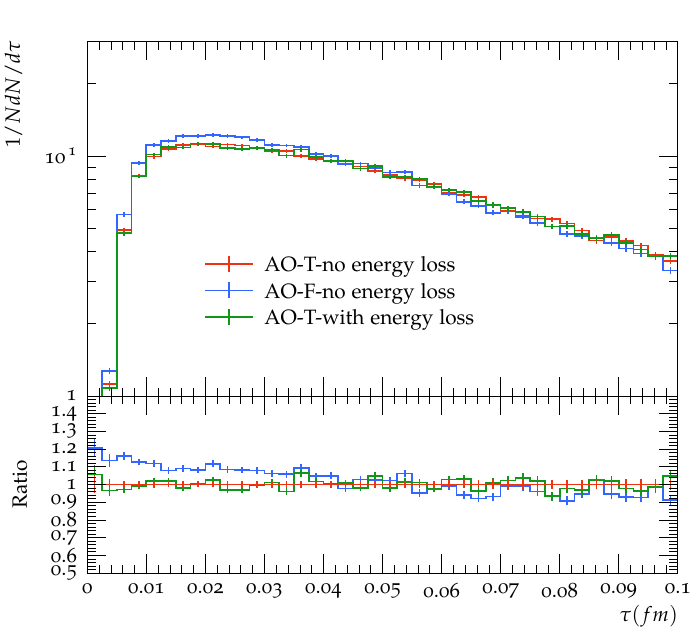}
    \label{fig:eloss-tau}
    }
    \label{fig:eloss}
    \caption{\subref{fig:eloss-raa} Jet $p_\perp$ spectra and \subref{fig:eloss-tau} reconstructed formation time distribution for the two angular ordering configurations without energy loss and for the configuration with energy loss and angular ordering turned on in the entire evolution.}
\end{figure}

Jet suppression as seen in \Fig{fig:raa} can also be due to parton energy loss. However, even in the case where both mechanisms give rise to similar nuclear modification factors, we can use the formation time distributions to distinguish between the two scenarios. To see that, we have generated events in which we require angular ordering for all splittings, but artificially make the initial parton lose energy by simply decreasing its momentum by 1\% as described in Section~\ref{sec:model}. \Fig{fig:eloss-raa} shows that the resulting suppression in the $p_\perp$ spectra is of the same order as the one in \Fig{fig:raa}, although the overall distribution is not exactly the same. At the same time, \Fig{fig:eloss-tau} shows that the reconstructed formation time is nearly identical to the distribution without energy loss and with angular ordering in the first splitting. This shows that the formation time distribution is sensitive only to the loss of angular ordering and not to a small amount of energy loss before the first splitting.

\section{Conclusions}

In this work we have explored a minimal scenario in which a single early re-scattering effectively breaks colour coherence in the parton shower, without introducing any explicit energy loss. Using controlled Monte Carlo simulations, we showed that disabling angular ordering in the first splitting leads to measurable modifications of the parton formation-time distribution in large-radius jets. The effect is attenuated but remains visible after full event reconstruction, demonstrating that the $\tau$-algorithm retains sensitivity to colour-coherence breaking.

The loss of angular ordering also induces an apparent suppression of the reconstructed jet yield, even though no energy is removed from the system. This shows that coherence effects alone can generate signatures similar to those usually attributed to medium-induced energy loss. By comparing this scenario to one with a small uniform reduction of the initial parton momentum, we demonstrated that the formation-time observable is insensitive to kinematical energy loss, confirming its role as a selective probe of coherence dynamics.

These results establish formation-time–based observables as a powerful tool to isolate and quantify colour-coherence effects in jet evolution. They suggest that future measurements, particularly in upcoming lighter-ion runs at the LHC, could exploit such observables to uncover subtle signatures of QGP droplets that would otherwise remain experimentally elusive. While a measurement of the formation time distribution along the lines discussed here is clearly very challenging, it could open up a possibility to observe for the first time a loss of colour coherence due to final state re-scattering.

\section*{Acknowledgments}

This study is part of a project that has received funding from the European Research Council (ERC) under the European Union's Horizon 2020 research and innovation programme  (Grant agreement No. 803183, collectiveQCD). This work is also supported by Fundação para a Ciância e a Tecnologia (FCT), under ERC-PT A-Projects ‘Unveiling’, financed by PRR, NextGenerationEU. LA acknowledge support by FCT under contract 2021.03209.CEECIND.

\bibliographystyle{unsrtnat}
\bibliography{refs}

\end{document}